\documentclass[fleqn,twoside]{article}
\usepackage{amssymb}
\usepackage{amsmath}
\usepackage{color}
\usepackage{epic}
\usepackage{espcrc2}
\usepackage{graphicx}
\usepackage{epsf}
\usepackage{axodraw}

\title{Long-distance effects and final state interactions in 
 $B\to \pi\pi K$ and $B\to K\overline KK$
decays}
 
\author{ A.~Furman$^{a}$,  R. Kami\'nski$^a$, L. Le\'sniak$^a$ and 
B. Loiseau$^b$\\
\addressmark{$^a$ Department of Theoretical Physics,\ The Henryk 
Niewodnicza\'nski Institute of Nuclear Physics,\\
\hspace{0.25cm}Polish Academy of Sciences, 31-342 Krak\'ow, Poland\\
$^b$ Laboratoire de Physique Nucl\'eaire et de Hautes 
\'Energies,\thanks{Unit\'e de Recherche des Universit\'es
Paris 6 et Paris 7, associ\'ee au CNRS} Groupe Th\'eorie,\\
\hspace{0.25cm}Univ. P. \& M. Curie, 4 Pl. Jussieu, F-75252 Paris, France \\
}}

\begin{document}

 \begin{abstract}
$B$ decays into $\pi\pi K$ and $K\overline KK$, where the $\pi\pi$ and 
$\overline KK$ pairs interact in isospin zero $S$-wave, are studied in 
the $\pi\pi$ effective mass range from threshold to 1.2 GeV.
The interplay of strong and weak decay amplitudes is analyzed using an 
unitary $\pi\pi$ and $K\overline K$ coupled channel model.
Final state interactions are described in terms of four scalar 
form factors constrained by unitarity and chiral 
perturbation theory. 
Branching ratios for the $B\to f_0(980)K$ decay, calculated 
in the factorization 
approximation with some QCD corrections, are too low as compared to recent 
data. 
In order to improve agreement with experiment,
we introduce long-distance contributions called charming penguins. 
Effective mass distributions, branching ratios and asymmetries are 
compared with the existing data from BaBar and Belle collaborations. 
A particularly large negative asymmetry in charged $B$ decays 
is predicted for one set of the charming penguin amplitudes.

 \end{abstract}

\maketitle
\section {Introduction}
Recent experimental results from $B$ factories indicate that 
charmless hadronic 
three-body decays are more frequent than two-body ones~\cite{eide04}.
Moreover one observes on Dalitz plots a definitive surplus of events at 
relatively small effective masses.
This is a signal of especially strong interactions between hadrons at 
not too high relative energies.
Many resonances are explicitly visible but in general the interference 
pattern is quite complicated.
Knowledge of these final state interactions is important to obtain 
a precise determination of the Cabibbo-Kobayashi-Maskawa (CKM) matrix 
elements.
Weak decay observables give  information on hadron-hadron 
interactions and internal quark or hadronic structure of the produced 
particles.

Prominent maxima in the $\pi^+\pi^-$ spectra are observed in the $B\to\pi^+\pi^-K$ 
decays in the $f_0(980)$ region 
\cite{bell02,aube04,baba04,baba03,baba0408,bell0412,bell0409,chen04}.
The $B^+\to f_0(980)K^+$ and 
$B^0\to f_0(980)K^0$ branching ratios are 
relatively large, and of the order of $10^{-5}$.
Direct and time-dependent $CP$-violating asymmetries are also measured.
The first Belle result on the $B^+\to f_0(980)K^+$ branching ratio 
\cite{bell02} has motivated the study of Chen \cite{chen03}.
In the perturbative QCD approach Chen finds that the non-strange content of the 
$f_0(980)$ can be important. 
According to Cheng and Yang \cite{cheng05}, subleading corrections 
due to intrinsic gluon effects inside $B$ meson may enhance the 
decay rate of $B\to f_0(980)K$. 
$B$~decays into scalar-pseudoscalar or scalar-vector particles have been 
studied by Minkowski and Ochs with a special emphasis on the presence of the 
lightest glueball \cite{mink04}.
The $\pi^+\pi^-$ mass spectrum in $B\to K\pi\pi$ decays, 
reported by Belle in 2003, is reproduced by a model amplitude of the coherent sum of 
$f_0(980)$, $f_0(1500)$ and a very broad glueball as a background.

In the present letter we study the $B$ decays into $\pi\pi K$ and $K\overline KK$.
We restrict ourselves to the case where the produced $\pi\pi$ or $K\overline K$ 
pairs interact in isospin zero $S$-wave from the $\pi\pi$ threshold to about 
$1.2$~GeV. 
One expects the $\pi\pi$ isospin two $S$-wave contribution to be small since the upper 
limit of the branching fraction for the $B^+\to \pi^+\pi^+K^-$ decay is less 
than $1.8\times 10^{-6}$~\cite{bababell}. 
Using the $K\overline K/\pi\eta$ branching ratio of $a_0(980)$~\cite{eide04} and 
the upper limit of $2.5\times 10^{-6}$~\cite{baba04a} for the branching ratio of 
$B^+\to a_0^0(980)K^+,\,a_0^0(980)\to \pi^0\eta$, 
one can estimate the branching fraction 
$\mathcal B (B^+\to a_0^0(980)K^+,\,a_0^0(980)\to K^+K^-)$ to be smaller than 
$1\times 10^{-6}$. 
This indicates that the $K\overline K$ isospin one $S$-wave amplitude is 
suppressed in the $B^\pm \to K^+K^-K^\pm$ decays. 

Two-pion $S$-wave rescattering effects have been recently considered by Gardner 
and Mei\ss ner~\cite{gard02}.
They study the effect of the $f_0(600)$ (or $\sigma$) resonance on the $B^0$ 
decay into $\pi^+\pi^-\pi^0$ in the range where the $\rho(770)\pi$ channel 
dominates.
The $\sigma\pi$ channel can play a~role in the determination of the CKM angle 
$\alpha$ from the $B^0\to\rho\pi$ decays.
Gardner and Mei\ss ner describe the broad $f_0(600)$ introducing a scalar form 
factor constrained by the chiral dynamics of low-energy meson-meson 
interactions~\cite{meis01}.
This scalar form factor is used instead of the commonly applied Breit-Wigner 
form to improve the description of the broad $\sigma$ and the understanding of 
the $B\to\rho\pi$ decays.

We extend the approach of ref.~\cite{gard02} to the $f_0(980)$ resonance.
The four strange and non-strange 
$\pi\pi$ and $K\overline K$ scalar form factors are 
constrained by chiral perturbation theory as developed by Mei\ss ner 
and Oller~\cite{meis01}.
Our final state interaction is, however, different from that of~\cite{gard02} 
and~\cite{meis01}.
Here we consider the unitary $\pi\pi$ and $K\overline K$ 
coupled channel model of~\cite{kami97}.

First the  $B\to (\pi\pi)_{S-wave}\,K$ and 
$B\to (K\overline K)_{S-wave}\,K$ decay amplitudes are calculated 
within the naive factorization 
approximation~\cite{baue87,ali98}.
Penguin amplitudes interfere destructively which leads to much too small 
$B\to f_0(980)K$ branching ratios.
Then we consider some QCD factorization corrections \cite{bene01} calculated by 
de~Groot, Cottingham and Whittingham \cite{groo03}.
These corrections are not sufficient to obtain agreement with experiment.
Further contributions are needed.
Here we include the long-distance contributions which have been considered in 
\cite{groo03} to improve their fit to hadronic charmless 
strange and non-strange two-body \mbox{$B$-decay} data.
These amplitudes, called charming penguin terms, originate from enhanced charm 
quark loops \cite{ciuc97}. 
They could, for instance, correspond to weak decays of $B$ to intermediate
$D_s^{(*)}D^{(*)}$ 
states followed by transitions to $f_0(980)K$ final states via $c\overline c$ annihilations. 
Their addition allows us to obtain a good agreement with the measured 
$B\to f_0(980)K$ branching fractions.

In sect.~2 we describe our weak decay amplitudes supplemented by the  scalar form factors.
Our model for the final state interactions is given in sect.~3.
Results of calculations and comparison with available data are presented in 
sect.~4. 
In sect.~5 we give some conclusions and final remarks.

\section{Amplitudes for the  \boldmath $B\to\pi\pi K$ and $B\to K\overline KK$ decays} 

We shall write the model amplitudes for the following decays: 
$B^\pm\!\to\!(\pi\pi)_SK^\pm$, $
B^\pm\to(K\overline K)_SK^\pm$, $B^0\!\to\!(\pi\pi)_SK^0,\ B^0\!\to\!(K\overline K)_SK^0$, 
$\overline B^0\!\to\!(\pi\pi)_S\overline K^0$ and 
$\overline B^0\!\to\!(K\overline K)_S\overline K^0$.
Here by $(\pi\pi)_S$ and $(K\overline K)_S$ we mean $\pi^+\pi^-$ or $\pi^0\pi^0$ 
and $K^+K^-$ or $K^0\overline K^0$ pairs in isospin zero $S$-wave.
\begin{center}
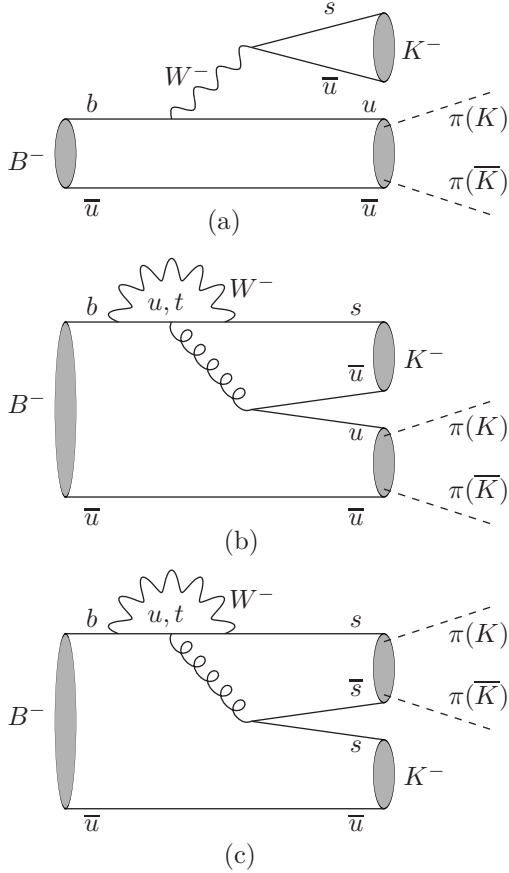
\begin{figure}[!h]
\begin{picture}(170,94)(-20,14)
\Text(3,55)[r]{$B^-$}
\Line(80,97)(130,110)
\Text(110,110)[b]{$s$}
\Line(80,97)(130,84)
\Text(110,83)[]{$\overline{u}$}
\GOval(130,97)(13,4)(360){0.7}
\Text(137,97)[l]{$K^-$}
\Photon(50,70)(80,97){3}{3.5}
\Text(65,87)[r]{$W^-$}
\GOval(10,57)(13,4)(360){0.7}
\Line(10,70)(130,70)
\Text(20,73)[b]{$b$}
\Text(125,73)[b]{$u$}
\Line(10,44)(130,44)
\GOval(130,57)(13,4)(360){0.7}
\DashLine(130,67)(170,80){3}
\DashLine(130,47)(170,34){3}
\Text(20,40)[t]{$\overline{u}$}
\Text(125,40)[t]{$\overline{u}$}
\Text(155,70)[l]{$\pi(K)$}
\Text(155,47)[l]{$\pi(\overline K)$}
\Text(70,26)[b]{(a)}
\end{picture}
\begin{picture}(140,115)(-20,5)
\Text(3,70)[r]{$B^-$}
\Line(10,100)(130,100)
\Text(20,103)[b]{$b$}
\Text(120,103)[b]{$s$}
\PhotonArc(50,100)(20,0,180){4}{6.5}
\Text(48,103)[b]{$u,t$}
\Text(72,110)[lb]{$W^-$}
\Gluon(80,67)(50,100){3}{5.5}
\GOval(130,87)(13,4)(360){0.7}
\Text(138,87)[l]{$K^-$}
\Line(80,67)(130,74)
\Line(80,67)(130,60)
\GOval(130,47)(13,4)(360){0.7}
\Text(120,60)[t]{$u$}
\Text(120,78)[b]{$\overline{u}$}
\Line(10,34)(130,34)
\GOval(10,67)(33,4)(360){0.7}
\DashLine(130,57)(170,70){3}
\DashLine(130,37)(170,24){3}
\Text(20,30)[t]{$\overline{u}$}
\Text(120,30)[t]{$\overline{u}$}
\Text(155,60)[l]{$\pi(K)$}
\Text(155,37)[l]{$\pi(\overline K)$}
\Text(70,17)[l]{(b)}
\end{picture}
\begin{picture}(140,106)(-20,26)
\Text(3,80)[r]{$B^-$}
\Line(10,110)(130,110)
\Text(20,113)[b]{$b$}
\Text(120,113)[b]{$s$}
\PhotonArc(50,110)(20,0,180){4}{6.5}
\Text(48,113)[b]{$u,t$}
\Text(72,120)[lb]{$W^-$}
\Gluon(80,77)(50,110){3}{5.5}
\GOval(130,97)(13,4)(360){0.7}
\DashLine(130,107)(170,120){3}
\DashLine(130,87)(170,74){3}
\Line(80,77)(130,84)
\Line(80,77)(130,70)
\GOval(130,57)(13,4)(360){0.7}
\Text(138,57)[l]{$K^-$}
\Text(120,70)[t]{$s$}
\Text(120,87)[b]{$\overline{s}$}
\Line(10,44)(130,44)
\GOval(10,77)(33,4)(360){0.7}
\Text(20,42)[t]{$\overline{u}$}
\Text(120,42)[t]{$\overline{u}$}
\Text(155,110)[l]{$\pi(K)$}
\Text(155,87)[l]{$\pi(\overline K)$}
\Text(70,26)[l]{(c)}
\end{picture}
\caption{Quark line diagrams for the $B^-$ decay: (a) tree diagram, 
(b) and (c) penguin diagrams. 
The 
spring-like lines represent gluon exchange and the dashed ones 
the $\pi\pi$ or 
$K\overline K$ isospin zero $S$-wave pairs.}
\end{figure}
\end{center}
\vspace{-0.8cm}

The possible quark line diagrams for the \mbox{$B^-$ decay}, together with the 
final state mesons, are shown 
in fig.~1.
For the $B^0$ decay there are only two types of penguin diagrams similar to 
those shown in figs.~1(b) and 1(c).
The tree diagram of fig.~1(a) is absent.
The $u\overline u$ or $s\overline s$ transitions into $\pi\pi$ or $K\overline K$ states, shown 
in fig.~1, are described by four scalar form factors.

In the approximation used in our approach the effective weak Hamiltonian 
$H$ is replaced by the sum of products of factorized currents 
\cite{ali98}.
We introduce some of the QCD factorization corrections and the 
charming penguin amplitudes considered in~\cite{groo03}. 
Then the $B^-\to(\pi^+\pi^-)_SK^-$ decay amplitude is
\begin{multline}
\label{eq:Heff}
      \langle(\pi^+\pi^-)_SK^-\vert H\vert B^-\rangle \\
     =\frac{G_F}{\sqrt{2}}\sqrt{\frac{2}{3}}\Big\{ \chi 
      \left[ P(m_{\pi\pi})U+C(m_{\pi\pi})\right] \Gamma_1^{n*}(m_{\pi\pi})+\\
  +
  \left[Q(m_{\pi\pi})V+\chi
  C(m_K)\right] \Gamma_1^{s*}(m_{\pi\pi})\Big\},
\end{multline}
where $G_F$ is the Fermi coupling constant and $\chi$ is a constant 
which will be estimated from the properties of the $f_0(980)$ decay. 
The functions $\Gamma_1^{n}(m_{\pi\pi})$ and 
$\Gamma_1^{s}(m_{\pi\pi})$ are the non-strange and strange pion scalar form 
factors depending on the effective pion-pion mass $m_{\pi\pi}$.
Furthermore the functions $P(m_{\pi\pi})$ and $Q(m_{\pi\pi})$ defined as
\begin{equation}
\label{eq:P}
P(m_{\pi\pi})=f_K (M_B^2-m_{\pi\pi}^2)F_0^{B\to(\pi\pi)_S}(M_K^2),
\end{equation}
\begin{equation}
\label{eq:Q}
Q(m_{\pi\pi})\!=\!\frac{2\sqrt{2}B_0}{m_b\!-\!m_s}(M_B^2\!-\!M_K^2)F_0^{B\to K}(m_{\pi\pi}^2),
\end{equation}
are proportional to the $B\!\to\!(\pi\pi)_S$ and $B\!\to\!K$ transition 
form factors, 
$F_0^{B\to(\pi\pi)_S}(M_K^2)$ and $F_0^{B\to K}(m_{\pi\pi}^2)$, respectively.
The masses of $B$ meson, kaon, pion, b-quark, strange-, down- and up-quarks are 
denoted by $M_B$, $M_K$, $m_\pi,\ m_b,\ m_s,\ m_d$ and $m_u$, respectively.
In eq.~(\ref{eq:P}) $f_K$ is the kaon decay constant.
In eq.~(\ref{eq:Q}) $B_0$ is related to the vacuum quark condensate: 
$B_0=-\langle 0\vert\overline qq\vert 0 \rangle/f_\pi^2$, $f_\pi$ being the pion decay 
constant equal to $92.4$ MeV.
We use the formula $B_0=m_\pi^2/(2 \widehat{m})$, where $\widehat{m}$ 
is the average mass of the light quarks 
$u$ and $d$. 
We put $\widehat{m}=5$~MeV and following ref.~\cite{ali98} we take
$m_s=0.122$~GeV and $m_b=4.88$~GeV.
The functions $U$ and $V$ in (\ref{eq:Heff}) depend on the combinations of the 
coefficients $a_i$ 
\cite{baue87,ali98,bene01,groo03} and on the products of the CKM matrix 
elements $\lambda_u=V_{ub}V_{us}^*$ and $\lambda_t=V_{tb}V_{ts}^*$:
\begin{equation}
\label{eq:U}
U\!\!=\!\lambda_u\!
\left[
a_1\!+\!a_4^{u}\!\!-\!a_4^c\!+\!\left(a_6^c\!-\!a_6^{u}\right)r
\right]
+\lambda_t\left(a_6^cr\!-\!a_4^c\right),
\end{equation}
\begin{equation}
\label{eq:V}
V=\lambda_u\left(a_6^c-a_6^{u}\right)+\lambda_ta_6^c,
\end{equation}
where the chiral factor $r=2 M_K^2/[(m_b+m_u)(m_s+m_u)]$.
In the numerical calculations we set $m_u=\widehat{m}$ and the 
coefficients $a_1$, $a_4^u$, $a_4^c$, $a_6^u$ and $a_6^c$, 
evaluated at the scale $\mu=2.1$~GeV, are taken from table III 
of~\cite{groo03}. These coefficients take into account some QCD factorization
corrections. We do not include small corrections  coming from hard gluon 
exchanges with spectator-quark, from annihilation terms and  from
electroweak penguin diagrams.
The charming penguin contribution can be parametrized as
\begin{equation}
\label{eq:C}
C(m)=\!-\!\left(M_B^2\!-\!m^2\right)\!f_\pi F_\pi\left(\lambda_uP_1^{GIM}\!\!+\!\lambda_tP_1\right),
\end{equation}
where $m$ is $m_{\pi\pi}$ or $M_K$, $F_\pi$ is the $B\to\pi$ transition 
form factor calculated at the zero $m_\pi^2$ limit and $P_1^{GIM}$, $P_1$ 
are complex parameters.
Determination of these parameters has been done in 
\cite{groo03,ciuc97,alek03,ciuc04} by fitting some charmless 
two-body \mbox{$B$-decay} data.

The neutral $\overline B^0\to(\pi^+\pi^-)_SK_S^0$ amplitude is similar 
to the charged one:
\begin{align}
\label{eq:Heff2}
  &\langle(\pi^+\!\pi^-)_SK^0_S\vert H\vert \overline B^0\rangle  
 = \frac{\langle(\pi^+\!\pi^-)_SK^-\vert H\vert B^-\rangle}
 {\sqrt{2}}\\
&{\rm with~} a_1=0 {\rm ~and~ } m_u\to m_d. \nonumber
\end{align}

The amplitudes for $B$-decays into three kaons read:
\begin{multline}
\label{eq:Heff3}
  \langle(K^+K^-)_SK^-\vert H\vert B^-\rangle  \\
  =\frac{G_F}{\sqrt{2}}\frac{1}{\sqrt{2}}\Big\{ \chi 
      \left[P(m_{K\overline K})U+C(m_{K\overline K})\right]\\
  \shoveright{\times \left[ \Gamma_2^{n*}(m_{K\overline K})+\Gamma_2^{n*}(\widetilde{m}_{K\overline K})\right] }\\
  +\left[Q(m_{K\overline K})V+\chi C(m_K)\right]  \\
  \times \left[ \Gamma_2^{s*}(m_{K\overline K})+\Gamma_2^{s*}(
  \widetilde{m}_{K\overline K})\right] \Big\}
 \end{multline}
 and
 \begin{align}
\label{eq:Heff4}
 &\langle(K^+\!\!K^-)_S\!K_S^0\vert H\vert \overline B^0\rangle  \!=\! 
 \frac{\langle (K^+\!\!K^-\!)_S\!K^-\!\vert H\vert\!  B^-\rangle\!}
 {\sqrt{2}} \\
&{\rm with~} a_1=0,\Gamma_2^{n,s*}(\widetilde{m}_{K\overline K})=0 {\rm ~and~ } m_u\to m_d.\nonumber
 \end{align}
Here $m_{K\overline K}$ is the $K^+K^-$ effective mass and $\widetilde{m}_{K\overline K}$ is 
the effective mass with the second possible $K^+K^-$ combination.
$\Gamma_2^{n}(m_{K\overline K})$ and $\Gamma_2^{s}(m_{K\overline K})$ are the non-strange 
and strange kaon scalar form factors.
Decay amplitudes with $(\pi^0\pi^0)_S$ and 
$(K^0\overline K^0)_S$ final states are 
the same as those with 
$(\pi^+\pi^-)_S$ and $(K^+K^-)_S$ pairs, respectively. 

If one uses the $a_i$ coefficients \cite{ali98} 
$a_4^{u}=a_4^c=a_4$ and $a_6^{u}=a_6^c=a_6$, then our formula (\ref{eq:Heff}) 
with $C(m)=0$ has the same algebraic structure as the $B^-\to\sigma\pi^-$ amplitude 
of ref.~\cite{gard02} (see their eq. (25)). 
The above equalities for $a_i$ are valid in naive factorization but not in
QCD factorization.
A particular feature of the two penguin contributions to the $b\to s$ 
transition is the near cancellation of these two terms in eq. (\ref{eq:U}) due 
to $a_4^c \approx a_6^c$ and $r\approx 1$.
This has been pointed out by Chernyak in his estimation of the scalar 
production in $B$ decays \cite{cher01} and also by Gardner and Mei\ss ner 
\cite{gard02}. 

Replacing the $\lambda_u$ and 
$\lambda_t$ values by their complex conjugate values $\lambda_u^*$ and 
$\lambda_t^*$ in 
eqs.~\eqref{eq:Heff}, \eqref{eq:Heff2}, \eqref{eq:Heff3} and 
\eqref{eq:Heff4} 
 gives the amplitudes for the $B^+\to (\pi^+ \pi^-)_S\,K^+$, 
$B^0\to (\pi^+ \pi^-)_S\,K^0_S$, 
$B^+\to (K^+ K^-)_S\,K^+$ and 
$B^0\to (K^+ K^-)_S\,K^0_S$ decays.

\section{Final state interactions}

In the $B$ decays considered above, one should include in the final states the 
$\pi\pi\to\pi\pi$ or the $K\overline K\to K\overline K$ rescattering and the 
$\pi\pi\to K\overline K$ or the $K\overline K\to\pi\pi$ transitions.
The $(\pi\pi)_S$ and $(K\overline K)_S$ pairs are formed from the 
$u\overline u,\ d\overline d$ and $s\overline s$ pairs.
The four scalar form factors appearing in eqs.~(\ref{eq:Heff}) and 
(\ref{eq:Heff3}) are defined \cite{meis01} as
\begin{eqnarray}
\label{eq:mat}
\begin{pmatrix}
\Gamma_1^n(m)\\
\Gamma_2^n(m)
\end{pmatrix} & = & \frac{1}{\sqrt{2}B_0}
\begin{pmatrix}
\langle 0\vert n\overline n\vert\pi\pi \rangle \\
\langle 0\vert n\overline n\vert K\overline K \rangle
\end{pmatrix}, \nonumber \\
\begin{pmatrix}
\Gamma_1^s(m)\\
\Gamma_2^s(m)
\end{pmatrix} & = & \frac{1}{\sqrt{2}B_0}
\begin{pmatrix}
\langle 0\vert s\overline s\vert\pi\pi \rangle \\
\langle 0\vert s\overline s\vert K\overline K \rangle
\end{pmatrix},
\end{eqnarray}
where $m$ is the effective $\pi\pi$ or $K\overline K$ mass, 
$n\overline n=(\overline uu+\overline dd)/\sqrt{2}$ and $\vert 0>$ denotes the vacuum state.
The final state interactions, which satisfy the unitarity 
constraints, are incorporated in the following formulae:
\begin{equation}
\label{eq:cont}
\begin{split}
\Gamma_i^{n,s}(m)&=R_i^{n,s}(m)\\
+&\sum_{j=1}^2\langle k_i \vert R_j^{n,s}(m)G_j(m)T_{ij}(m) \vert k_j \rangle,
\end{split}
\end{equation}
where $\vert k_i \rangle$ and $\vert k_j \rangle$ represent 
the wave functions of two mesons in the momentum space and the indices 
$i,j=1,2$ refer to the $\pi\pi$ and $K\overline K$ channels, respectively.
The center of mass channel momenta are given by 
$k_1=\sqrt{m^2/4-m_\pi^2}$ and $k_2=\sqrt{m^2/4-m_K^2}$.
The matrix $T$ is the two-body scattering matrix.
Here we  use the solution $A$ of the 
$\pi\pi$ and $K\overline K$ coupled channel model~\cite{kami97}.
As we restrict ourselves to 
$m_{\pi\pi}\lesssim 1.2$~GeV, the third effective 
$(2\pi)(2\pi)$ coupled channel considered in this model, with 
a threshold around $1.4$~GeV, has a small effect.
The functions $G_i(m)$ are the free Green's functions defined 
in~\cite{kami97} and $R_i^{n,s}(m)$ are the production functions 
responsible for the initial 
formation of the meson pairs prior to rescattering.
The production functions have been derived by Mei\ss ner 
and Oller in the one-loop approximation of the chiral perturbation theory~\cite{meis01}. 
Using their eqs. (37), (38), (41) and (44) one obtains 
\begin{equation}\label{eq:R}
\begin{split}
R^n_1(m)=& &\!\!\!\!\!0.566+0.414\,m^2,\\      
R^n_2(m)=&-&\!\!\!\!\!0.322+0.527\,m^2,\\      
R^s_1(m)=&-&\!\!\!\!\!0.036+0.353\,m^2,\\      
R^s_2(m)=& &\!\!\!\!\!0.071+0.338\,m^2,        
\end{split}
\end{equation}
where $m$ is in GeV.

If one considers only the on-shell contributions in 
eq.~\eqref{eq:cont} then the scalar form factors can be written in 
terms of the phase shifts $\delta_{\pi\pi}(m)$, $\delta_{KK}(m)$ and of the 
inelasticities $\eta(m)$:
\begin{multline}
\label{eq:gam1}
\Gamma_1^{n,s*}(m) \\
=\frac{1}{2}\bigg[
R_1^{n,s}(m)\left(
1+\eta(m)e^{2i\delta_{\pi\pi}(m)}\right)-iR_2^{n,s}(m)    \\
 \times\sqrt{\frac{k_2}{k_1}}\sqrt{1-\eta^2(m)}
e^{i[\delta_{\pi\pi}(m)+\delta_{K\overline K}(m)]}\bigg],
\end{multline}
\begin{multline}
\label{eq:gam2}
\Gamma_2^{n,s*}(m) \\
=\frac{1}{2}\bigg[
R_2^{n,s}(m)\left(
1+\eta(m)e^{2i\delta_{K\overline K}(m)}\right)-iR_1^{n,s}(m) \\
 \times\sqrt{\frac{k_1}{k_2}}\sqrt{1-\eta^2(m)}
e^{i[\delta_{\pi\pi}(m)+\delta_{K\overline K}(m)]}\bigg].
\end{multline}
Below the $K\overline K$ threshold $\eta(m)=1$ and
\begin{equation}
\Gamma_1^{n,s*}(m)  =  R_1^{n,s}(m)\cos\delta_{\pi\pi}(m)\ e^{i\delta_{\pi\pi}(m)}.
\label{ga1nsbt}
\end{equation}
Particularly interesting is the $m_{\pi\pi}$ range where 
the phase shifts $\delta_{\pi\pi}$ are close to $180^\circ$. 
Then one expects a {\it maximum} in 
$\vert \Gamma_1^{n,\,s}\vert$.
As we shall see in the next section, this is the case for the production 
of the $f_0(980)$ resonance. Note that the $\Gamma_1^{n,\,s}$ 
is zero when $\delta_{\pi\pi}=\pi/2$.

The off-shell contributions to the form factors are very much 
model dependent and are not considered in the following 
calculations.

\section{Results}

The amplitudes for the $B \to (\pi\pi)_S K$ decays considered in sect.~2
depend only on the effective mass $m_{\pi\pi}$. 
Integrating on the Dalitz plot over the kinematically allowed range of  $m_{\pi K}$, 
one obtains the differential $B \to \pi\pi K$ decay distribution

\begin{equation}
\frac{d\Gamma}{dm_{\pi\pi}} = 
\cfrac{m_{\pi\pi}\,k_1\,p_K}
{4\,M_B^3\,(2\pi)^3} 
\times |\mathcal{M}(B \to (\pi\pi)_S K)|^2,
\label{eq:dgpi}
\end{equation}
where  
$k_1$ and 
 $p_K= \sqrt{E_K^2(m_{\pi\pi})\,-\,m_K^2} $
are the pion and kaon momenta in the $\pi\pi$ center of mass system, 
$E_K(m_{\pi\pi})\, = \,\frac{1}{2}(M_B^2-m_{\pi\pi}^2-m_K^2)/m_{\pi\pi}$ 
being the corresponding 
kaon energy.
In \eqref{eq:dgpi} $\mathcal{M}$ denotes the decay amplitude given by 
eq. \eqref{eq:Heff} or \eqref{eq:Heff2}. 
Dividing $d\Gamma/dm_{\pi\pi}$ by the appropriate $B^+$ or $B^0$ total width
$\Gamma_B$ one obtains the differential branching ratio 
$d\mathcal{B}/dm_{\pi\pi}$.

Before presenting our results we fix the constants which appear in the
formulae for the decay amplitudes \eqref{eq:Heff}, \eqref{eq:Heff2}, 
\eqref{eq:Heff3} and \eqref{eq:Heff4}. 
The masses of pions, kaons, $B$-mesons and their life times, 
the values of the Fermi coupling constant $G_F$ and the kaon decay 
constant $f_K = 0.1598$ GeV
are taken from \cite{eide04}. 
For the kaon mass we use the average of the charged and neutral ones. 
Following
the results obtained in \cite{gatt00} and applied in \cite{gard02} 
we use $F_0^{B\to(\pi\pi)_S}(M_K^2)$=0.46 for
 the  $B \to (\pi\pi)_S$ transition form factor. 

The $B \to K$ transition form factor is approximated by a constant 
equal to 0.39 which is close
to the number $F_0^{B\to K}(0) = 0.379$ quoted by 
Bauer, Stech and Wirbel in table 14 of~\cite{baue87}.
This approximation is justified since we consider a relatively narrow range of the
effective masses $m_{\pi\pi}$. These masses are much smaller than the mass of the 
heavy excited
$B$ meson used in polar models of the transition form factor.
So we fix $F_0^{B\to K}(m_{\pi\pi}^2) \approx F_0^{B\to K}(m_{f_0}^2)$ where
 $m_{f_0}$ is the $f_0(980)$ mass.  The value 0.379 quoted in \cite{baue87} as 
 well as a more recent value of 0.33, obtained by Ball and Zwicky in 
 \cite{Ball05},
 are within the limits given by Beneke and Neubert in table 1 of \cite{Beneke03}
 for  $F_0^{B\to K}(0)= 0.34\pm0.05$.

Using the Wolfenstein representation \cite{wolf83}, the CKM matrix elements are
written in a form accurate to the level of $\lambda^6$ \cite{flei02}.
The values of the parameters, taken from the CKMfitter Group, are: 
$\lambda \equiv V_{us} = 0.2265$, $\overline \rho = 0.189$, 
$\overline \eta = 0.358$ and $A = 0.801$~\cite{char04}.

The constant $\chi$ will be fitted to the experimental branching ratio of the
decay $B^+ \to f_0(980) K^+$.
Once fixed it will be used to make absolute  model predictions in the whole 
$m_{\pi\pi}$ range studied here for the
$B^\pm \to \pi^+\pi^- K^\pm$ reaction and for other decays like 
$B^0 \to \pi^+\pi^- K^0_S$, $B^+ \to K^+ K^- K^+$ and $B^0 \to K^+ K^- K^0_S$.
The value of $\chi$ is, however, not arbitrary. 
It can be estimated from the following considerations.
First we shall concentrate ourselves on the $m_{\pi\pi}$ range close to the relatively narrow
resonance $f_0(980)$ clearly visible in the $B$ decays into $\pi^+\pi^-K$.
The $f_0(980)$ decays mainly into $\pi\pi$.
The coupling constant of $f_0(980)$ to the $\pi\pi$ pair can be 
approximated by 

\begin{equation}
g_{f_0\pi\pi} = m_{f_0}\,\sqrt{\frac{8\,\pi\,\Gamma(f_0\to\pi\pi)}{k_1(m_{f_0})}}
\label{eq:gfpi}
\end{equation}
(see, for example, eq.~(36) of~\cite{kami99}).
Here $\Gamma(f_0\to\pi\pi)$ is the $\pi\pi$ partial
width of the $f_0(980)$ and $k_1(m_{f_0})$ is the 
pion momentum in the $f_0(980)$ rest frame. 
The two scalar form factors $\Gamma_1^n(m_{\pi\pi})$ and 
$\Gamma_1^s(m_{\pi\pi})$ are strongly
peaked at the $f_0(980)$ mass; one can show, however, using 
eqs.~\eqref{eq:R} and~\eqref{eq:gam1} that 
 $\vert \Gamma_1^n(m_{f_0})\vert \gg \vert\Gamma_1^s(m_{f_0})\vert$.
Therefore one can write the approximate relation 

\begin{equation}
\chi = \frac{g_{f_0\pi\pi}}{m_{f_0}\,\Gamma_{tot}(f_0)}\frac{1}
{|\Gamma_1^n(m_{f_0})|} 
\label{eq:chi}
\end{equation}
(see, for instance, eq.~(35) of~\cite{gard02} for the $f_0(600)$ case).
Here $\Gamma_{tot}(f_0)$ is the total $f_0(980)$ width. 
In ref. \cite{kami99} 
$\Gamma_{tot}(f_0)$=($71\pm 14$) MeV.
For $m_{f_0} = 0.98$ GeV we obtain from eq.~(\ref{eq:gam1}) 
$|\Gamma_1^n(m_{f_0})| \approx 0.96$.
If 
$\Gamma_{tot}(f_0) \approx \Gamma(f_0 \to \pi\pi) = 60$ MeV 
then 
eqs.~\eqref{eq:gfpi} and~\eqref{eq:chi} give
$\chi \approx 30$ GeV$^{-1}$.

In this paper we do not attempt to make our own fits to data by 
adjusting the constants $P_1^{GIM}$ and $P_1$ in the function $C(m)$ 
of~\eqref{eq:C}. 
We present the results obtained with the 
charming penguin amplitudes determined in~\cite{groo03} 
and~\cite{ciuc04}.
 The theoretical curves shown in 
figs.~\ref{fig:babarpm} to~\ref{fig:belle0} 
correspond to the first set of amplitudes.

\subsection{\boldmath $B^\pm \to \pi^+\pi^-K^\pm$ decays}

\begin{figure}[!ht]
\includegraphics*[width=15pc]{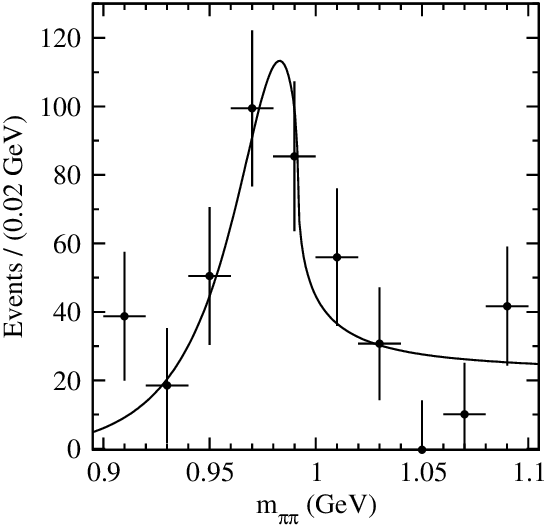}
\caption{Effective $\pi^+\pi^-$ mass distribution in 
$B^\pm\to \pi^+\pi^-K^\pm$ decays. The BaBar data are taken 
from~{\protect \cite{baba03}}. The solid line results from our 
model.} 
\label{fig:babarpm}
\end{figure}

\begin{figure}[!ht]
\includegraphics*[width=18pc]{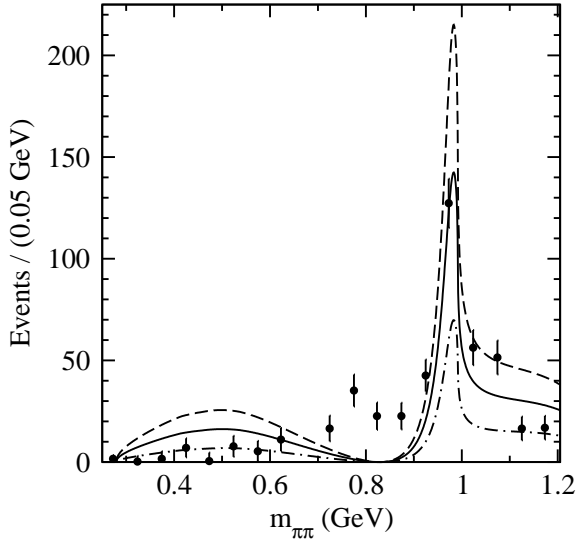}
\caption{Comparison of the Belle~\cite{bell0412} $\pi^+\pi^-$ effective 
mass distribution for the $B^\pm\to \pi^+\pi^-K^\pm$ decays 
with our model. The dashed line corresponds to the 
$B^+\to \pi^+\pi^-K^+$ decays, the dotted-dashed line is for the 
$B^-\to \pi^+\pi^-K^-$ decays and the solid line is the average for the 
$B^+$ and $B^-$ decays.}
\label{fig:bellepm}
\end{figure}

In figs.~\ref{fig:babarpm} and~\ref{fig:bellepm} we show a comparison 
of the $\pi\pi$ effective mass distributions for the 
$B^\pm  \to \pi^+ \pi^- K^\pm$ decays
calculated in our model with the results obtained by 
BaBar~\cite{baba03} and Belle~\cite{bell0412}.

The branching fraction 
$\mathcal{B}(B^\pm\to f_0(980)K^\pm,\,f_0(980)\to \pi^+\pi^-)=(9.2\pm 1.2^{+2.1}_{-2.6})\times 10^{-6}$ 
obtained by BaBar~\cite{baba03} can be reproduced in our model for the 
value of $\chi=35.0\rm ~GeV^{-1}$. This value is close to our 
estimation given above. 
The Belle Collaboration has reported a slightly smaller 
value of the branching ratio 
$(7.55\pm 1.24\pm 0.69^{+1.48}_{-0.96})\times 10^{-6}$~\cite{bell0412}. 
Both values are compatible within their error bars. The average value 
given by the Heavy Flavor Averaging Group (HFAG) is equal to 
$(8.49^{+1.35}_{-1.26})\times 10^{-6}$~\cite{hfag}. 
For this value the constant $\chi$ is 
$33.5\rm ~GeV^{-1}$ when we use the charming penguin amplitudes 
of~\cite{groo03}. 

In fig.~\ref{fig:babarpm} the normalization of the theoretical curve 
to the data in the $f_0(980)$ range
is based on the total number of events seen by BaBar for 
$m_{\pi\pi}$ from 0.9 to 1.1 GeV, multiplied by a correction factor 
of 0.92 being a ratio of the branching ratios $8.49\times 10^{-6}$
(HFAG's value) and 
$9.2\times 10^{-6}$ (BaBar's value).

In fig.~\ref{fig:bellepm} the solid curve is normalized at 
$m_{\pi\pi}=976\rm ~MeV$. This corresponds to the maximum of the 
background subtracted mass 
distribution. We calculate it from fig. 9e 
of~\cite{bell0412} as $133$~events/50~MeV. 
This number is obtained by taking into account two factors. 
The first factor, equal to 92\%, follows from the fraction of the 
$f_0(980)$ components in the full spectrum resulting from the 
solution 1 of the Belle model, called $K\pi\pi -C_0$~\cite{bell0412}. 
The second factor, equal to 1.125, comes from the ratio of the above 
mentioned branching ratios: $8.49\times 10^{-6}$ and 
$7.55\times 10^{-6}$. 

One can see from fig.~\ref{fig:babarpm} that our model describes quite well the $\pi^+\pi^-$
spectrum measured by BaBar in vicinity of $f_0(980)$. 
The model depicts also a very pronounced maximum seen by Belle near 1 GeV (fig.~\ref{fig:bellepm}). 
This maximum is attributed to the $f_0(980)$ resonance.
At lower $\pi\pi$ masses near 500 MeV one can notice a broad theoretical  maximum which we can
relate to the $\sigma$ or $f_0(600)$ meson. 
Experimental data in this mass range are not in disagreement with this feature of our model although the
present errors are too large to draw a definite conclusion supporting the evidence of $f_0(600)$ in the 
$B^\pm \to \pi^+\pi^- K^\pm$ decay.
Also the preliminary data of the BaBar Collaboration \cite{baba0408} 
below $600$~MeV show some enhancement of the $\pi^+\pi^-K^\pm$ events 
over the background. 
Let us remark that both collaborations have not included the $\sigma$ 
meson in their fits to data. 
Integrating the $\pi\pi$ spectrum between the $\pi^+\pi^-$ threshold and 700 MeV we find the average
branching ratio for the $B^+ \to \sigma K^+$ and $B^- \to \sigma K^-$ 
decays equal to $3.9\times 10^{-6}$. 
A~surplus of events near $0.8$~GeV is due to the $B^\pm\to \rho^0K^\pm$ 
decay which are not taken into account in this model since we concentrate ourselves 
on the \mbox{$S$-wave} $\pi^+\pi^-$ events. 

In figs. \ref{fig:babarpm} and \ref{fig:bellepm} we have 
shown the averaged $\pi^+\pi^-$ spectra of the two decays: 
$B^+ \to \pi^+\pi^-K^+$ and 
$B^- \to \pi^+\pi^-K^-$. They have been calculated for 
the charming penguin amplitudes $C(m)$ of eq.~\eqref{eq:C} fitted in 
ref.~\cite{groo03} by $P_1=(0.068\pm 0.007)\exp[i(1.32\pm 0.10)]$ and
$P_1^{GIM}=(0.32\pm 0.14)\exp[i(1.0\pm 0.27)]$. 
Note that the strength of the $P_1$ contribution to $C(m)$ 
is about ten times larger than that of $P_1^{GIM}$. 
If we neglect both of them the 
branching ratios are smaller by a~factor of about 4. 

Using these long distance contributions we have found a very pronounced direct $CP$ asymmetry in the $\pi^+\pi^-$ spectra.
There are many more decays of $B^+$ into $\pi^+\pi^-K^+$ than of 
$B^-$ into $\pi^+\pi^-K^-$ (see fig.~\ref{fig:bellepm})
This asymmetry for the above choice
of parameters is even higher than the direct $CP$ asymmetry recently found in the $B^0 \to K^+ \pi^-$ and 
$\overline B^0 \to K^- \pi^+$ decays \cite{baba04pik,chao04}. 
The charge asymmetry is defined as 
\begin{equation}
\mathcal{A}_{CP} = \frac{\frac{d\,\Gamma(B^- \!\to \pi^{\!+}\pi^{\!-}\!K^-)}
{d\,m_{\pi\pi}}\!-\!
\frac{d\,\Gamma(B^+ \!\to \pi^{\!+}\pi^{\!-}\!K^+)}{d\,m_{\pi\pi}}}
{\frac{d\,\Gamma(B^- \!\to \pi^{\!+}\pi^{\!-}\!K^-)}{d\,m_{\pi\pi}}\!+
\frac{d\,\Gamma(B^+ \!\to \pi^{\!+}\pi^{\!-}\!K^+)}{d\,m_{\pi\pi}}}.
\label{eq:aCP}
\end{equation}
If we integrate $d\Gamma/dm_{\pi\pi}$ between $0.9$ and $1.1$ GeV 
then $\mathcal{A}_{CP}=-0.52\pm 0.12$.
The errors come from the 
uncertainties of the charming penguin amplitudes determined 
in~\cite{groo03}. 
It would be very useful to confront this number with a future 
experimental determination of this
asymmetry in the $B^\pm \to f_0(980) K^\pm$ decays. 

The charge $CP$ violating asymmetry is very 
sensitive to the magnitude and the phase of
the charming penguin contribution.
Using the different approach of~\cite{ciuc97} and the fits 
presented in~\cite{ciuc04} for the
$B \to K\pi$ decays 
the value 
$P_1 =(0.08\pm 0.02)\exp[-i(0.6\pm 0.5)]$ has been obtained, while 
 $P_1^{GIM}$ has not been 
determined.
With this value of $P_1$ and with $P_1^{GIM}=0$, one obtains a~good agreement 
with the HFAG branching fraction for $\chi=23.5~{\rm GeV}^{-1}$. 
Then the charge asymmetry is positive and equal to 
$0.20\pm 0.20$.
We see that the different charming 
penguin amplitudes fitted to data are not in mutual agreement. 
However, the two analyses 
are based on different data sets.
In~\cite{groo03} 18 different branching ratios have been fitted and
in~\cite{ciuc04} the fit has been performed on 8 observables for the
$B \to K\pi$ decays. 

In the following subsections the predictions for the two penguin 
amplitudes are given without any readjustment of the 
constants $\chi$.

\subsection{\boldmath $B^0 \to \pi^+\pi^- K^0$ decays}

\begin{figure}[!ht]
\includegraphics*[width=15pc]{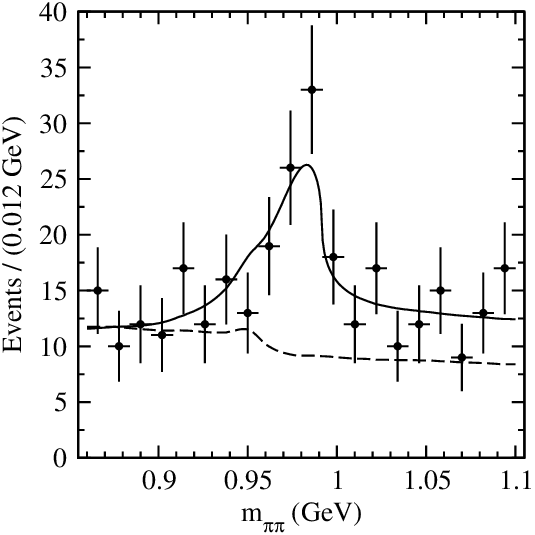}
\caption{Effective $\pi^+\pi^-$ mass distribution 
  for the 
$B^0 \to \pi^+\pi^- K^0$ decays. The BaBar data and the dashed line 
corresponding to the background, are taken from \cite{baba04}. The 
solid line is our model result.}
\label{fig:babar0}
\end{figure}

In fig.~\ref{fig:babar0} we show the comparison of the model predictions 
for the neutral $B$ decays with the BaBar results \cite{baba04}.
Here the experimental background contribution is added to the 
theoretical part calculated from eq. \eqref{eq:Heff2}. 
The average branching ratio for the $B^0$ and $\overline B^0$ decays 
into $(\pi^+\pi^-)_SK_S^0$, for
the $\pi^+\pi^-$ mass between 0.85 and 1.1 GeV, equals to 
$2.93\times 10^{-6}$.
Twice this value compares well with the experimental determination of
the BaBar Collaboration 
$\mathcal{B}(B^0 \to f_0(980) K^0) \times \mathcal{B}(f_0(980) \to \pi^+\pi^-) = 
(6.0\pm 0.9 \pm 0.6 \pm 1.2)\times10^{-6}$. 
Our curve in fig.~\ref{fig:babar0} is normalized to the total number of 
events multiplied by the ratio of $2\times 2.93/6.0=0.98$.

The branching ratio diminishes by a~factor of 18 
if the charming penguins amplitudes are omitted. 
This was expected due to the near cancellation, mentioned in 
section 2, between the two penguin diagram contributions. The absence 
of tree diagram in neutral $B$~decays explains the difference with the charged $B$~decays 
where the branching ratio drops by a~factor of 4 if the 
charming penguin terms are not present.

The direct $CP$ violation asymmetry between the decay of $B^0$ and 
$\overline B^0$ into $(\pi^+\pi^-)_S K_S^0$ defined as
\begin{equation}
\mathcal{A} = 
\cfrac{\mathcal{B}(\overline B^0 \!\!\!\!\to \!\pi^+\pi^- 
\overline K^0) \!- \!
\mathcal{B}(B^0 \!\!\to\! \pi^+\pi^- K^0)}
{\mathcal{B}(\overline B^0 \!\!\!\!\to\! \pi^+\pi^- \overline K^0) \!+\! 
\mathcal{B}(B^0 \!\!\to\! \pi^+\pi^- K^0)}
\label{a0}
\end{equation}
is much smaller than the asymmetry \eqref{eq:aCP} for the 
$B^\pm \to (\pi^+\pi^-)_SK^\pm$ decays.
It amounts to $0.01\pm 0.10$ when 
calculated with the charming penguin parameters of~\cite{groo03} in the $m_{\pi\pi}$ 
range between 0.85 and 1.1 GeV. 
The reason of its smallness is due to the absence of the tree diagram 
contribution (fig. 1a) for the $B^0$ or $\overline B^0$ decays.

The BaBar~\cite{aube04} and Belle~\cite{bell0409} values for this 
asymmetry,
$\mathcal{A}= 0.24 \pm 0.31 \pm 0.15$ and
$\mathcal{A}=-0.39\pm 0.27\pm 0.08$, respectively, 
have large experimental errors and agree with our result.
The HFAG average is $-0.14\pm 0.22$~\cite{hfag}.
\begin{figure}[!ht]
\includegraphics*[width=15pc]{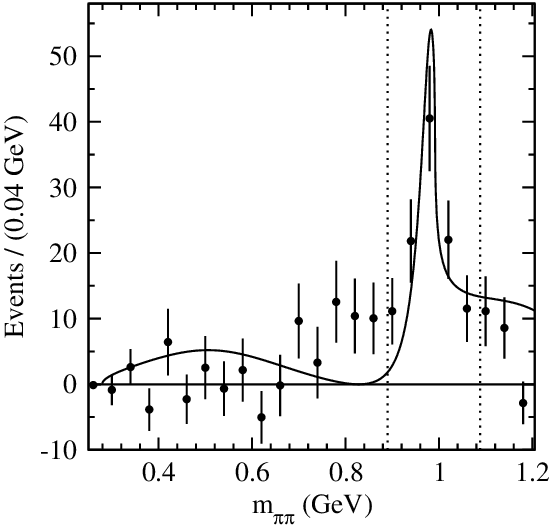}
\caption{Comparison of the Belle data~\cite{chen04} (after a background 
subtraction) with our model (solid line) for the 
$B^0\to \pi^+\pi^-K^0_S$ decays. 
Dotted vertical lines delimit a band of the $f_0(980)$ events 
used in the curve normalization.}
\label{fig:belle0}
\end{figure} 
In fig.~\ref{fig:belle0} we compare the $\pi^+\pi^-$ spectrum presented 
by K.-F. Chen on behalf of the Belle Collaboration for the 
$B^0 \to \pi^+\pi^-K^0_S$ decays~\cite{chen04} 
with our calculation when the background is subtracted.
Our curve is normalized to the number of $94$ events attributed to the 
$f_0(980)\,K^0_S$ decay in the $m_{\pi\pi}$ range between 0.89 and 
1.088 GeV~\cite{bell0409}.
The Belle experimental determination of the branching ratio for the 
$B^0 \to \pi^+ \pi^- K^0_S$ decay is not yet available.
As in the $B^\pm \to \pi^+\pi^- K^\pm$ 
case, shown in fig.~\ref{fig:bellepm}, we predict some $\sigma$ 
contribution of the decay amplitude in the low 
$\pi^+\pi^-$ range below the position of $\rho(770)$ enhancement
visible in the data. 
The zero value of our spectrum near 
$m_{\pi \pi}$=0.8 GeV, also present in fig.~3, comes from 
$\delta_{\pi \pi} = \pi/2$ 
(see eq.~\eqref{ga1nsbt}). 

We have also calculated the time-dependent $CP$ asymmetry in the 
neutral $B^0$ decays into $\pi^+\pi^-K^0_S$:

\begin{equation}
A(t) = 
\cfrac{\frac{d\Gamma(\overline B^0 \to \pi^+\pi^- K_S^0)}{dt} - 
\frac{d\Gamma(B^0 \to \pi^+\pi^- K_S^0)}{dt}}
{\frac{d\Gamma(\overline B^0 \to \pi^+\pi^- K_S^0)}{dt} + 
\frac{d\Gamma(B^0 \to \pi^+\pi^- K_S^0)}{dt}}.
\label{afCP}
\end{equation}
The time dependence of this asymmetry can be approximated as 
\begin{equation}
A(t) = \mathcal S\, sin(\Delta m t)+\mathcal{A}\, cos(\Delta m t),
\label{afCPapp}
\end{equation}
where $\Delta m$ is the difference between the masses of the heavy and 
light $B$ meson eigenstates. 

In the $f_0(980)$ mass range between 0.85 and 1.1 GeV the asymmetry 
parameter $\mathcal S$ equals to $-0.63\pm 0.09$. This result corresponds to the 
charming penguin parameters of~\cite{groo03}. 
The BaBar result is $\mathcal S = -0.95^{+0.32}_{-0.23} \pm 0.10$ \cite{aube04} 
and the Belle number 
$\mathcal S = +0.47 \pm 0.41 \pm 0.08$ \cite{bell0409}.
The two experimental results are not in agreement with each other 
but their experimental errors are large.
Our result agrees better with the BaBar value. 
The HFAG~\cite{hfag} gives the average 
$\mathcal S = -0.39 \pm 0.26$ which is in agreement with our 
prediction of $-0.63$. 
The value of $\mathcal{S}$ for the 
charming penguin amplitudes of~\cite{ciuc04}, considered in 
section 4.1, is $-0.77$. 

\subsection{\boldmath $B \to K\overline K K$ decays}

We have calculated the $(K^+K^-)_S$ spectra and the branching ratios 
for the $B^+ \to (K^+K^-)_S K^+$ and 
$B^- \to (K^+K^-)_S K^-$ decays. 
For the charming penguin amplitudes of~\cite{groo03} one 
obtains a~large direct $CP$ violating 
asymmetry of $-0.44\pm 0.12$ 
in the $K^+K^-$ mass range between the threshold and $1.1$ GeV.
The average branching ratio for the above mass range equals to 
$(1.8\pm 0.4)\times 10^{-6}$.
This value is below the upper limit of $2.9\times 10^{-6}$ found by 
Garmash {\it et al.} for the branching fraction 
$\mathcal B(B^+ \to f_0(980)K^+,
f_0(980) \to K^+K^-)$~\cite{bell0412}.
The theoretical $(K^+K^-)_S$ spectrum is flat in the range between 
$1.0$ and $1.2$ GeV and agrees well with the experimental 
distribution shown in fig. 13d of~\cite{bell0412}. 
As in the case of the $B^\pm \to \pi^+\pi^-K^\pm$ decays the 
asymmetry for the $B^\pm \to (K^+K^-)_SK^\pm$ process strongly depends 
on the charming penguin amplitude. 
With the amplitudes of~\cite{ciuc04} considered in 
section 4.1, the 
asymmetry is positive and 
equal to $0.29\pm 0.21$, the average branching ratio being 
$(1.7\pm 0.7)\times 10^{-6}$.
In these calculations we neglect the symmetrized form factors 
$\Gamma_2^{n,s}(\widetilde m_{K\overline K})$ in eq. (8) assuming that 
for small masses of $m_{K\bar K}$ 
the $\widetilde m_{K\overline K}$ masses are sufficiently high and 
that the $\Gamma_2$ form factors decrease
rapidly with increasing $\widetilde m_{K\overline K}$.

For the $B^0 \to (K^+K^-)_S K^0_S$ and 
$\overline B^0 \to (K^+K^-)_S K^0_S$ decays we have obtained very 
small direct $CP$ violating asymmetries
$\mathcal A = 0.01\pm 0.10$ and $\mathcal A = 0.001\pm 0.001$ for the penguin amplitudes 
of~\cite{groo03} and~\cite{ciuc04}, respectively.
These numbers agree well with the experimental findings of 
Belle~\cite{bell0409} $(-0.08 \pm 0.12 \pm 0.07)$ and 
BaBar~\cite{baba0408076} $(-0.10 \pm 0.14 \pm 0.04)$. 
The parameter $\mathcal S$, equal to $-0.64$ or $-0.77$, depending 
on the set of penguin amplitudes, is also in general 
agreement with the results of $-0.74 \pm 0.27 ^{+0.19}_{-0.39}$ and 
$-0.55 \pm 0.22 \pm 0.04 \pm 0.11$ reported by Belle~\cite{bell0409} 
and BaBar~\cite{baba0408076}, respectively.

\begin{table*}[t]
\begin{center}
\caption{Average branching fractions $\mathcal{B}$ in units of 
$10^{-6}$, asymmetries $\mathcal{A}_{CP},\ \mathcal{A}$ and 
$\mathcal{S}$ of our model compared to the average  values of 
HFAG~\protect{\cite{hfag}}. The $m_{\pi\pi}$ mass ranges for the
 $B^\pm\to f_0(980)K^\pm$ and for the $B^0\to f_0(980) K^0$ decays
 are $(0.9,1.1)$ GeV and $(0.85,1.1)$ GeV, respectively. The upper limit of the
 $\left(K^+K^-\right)_S$ or $\left(K^0_SK^0_S\right)_S$ effective mass is $1.1$
 GeV. 
 The model errors come from the uncertainties of 
the charming penguin amplitudes $C(m)$~(eq. (\protect{\ref{eq:C}})) 
determined in the fits of~\protect{\cite{groo03}} (model I) or 
\protect{\cite{ciuc04}} (model II). The experimental errors for $\mathcal S$ 
in the $B^0\to (K^+K^-)_SK^0_S$ decay are the statistical ones.}
\medskip
\begin{tabular}{ccccc}
\hline
	   &            & Average                      &  Model I       & Model II\\

\raisebox{1.5ex}[0pt][0pt]{$B$ decay mode} &  &  HFAG's values 
	   &$\chi=33.5$ GeV$^{-1}$& $\chi=23.5$ GeV$^{-1}$\\
\hline
\hline\\[-11pt]
 \raisebox{-1.5ex}[0pt][0pt]{$B^\pm\to f_0(980)K^\pm,~f_0\to \pi^+\pi^-$}    &$\mathcal{B}$   & $8.49^{+1.35}_{-1.26}$  & 8.49 (fit)  & 8.46 (fit) \\[2pt]
                                          &$\mathcal{A}_{CP}$& no data       &$-0.52\pm 0.12$ & $0.20\pm 0.20$\\
\hline\\[-11pt]
                                          &$\mathcal{B}$     & $6.0\pm{1.6}$ &$5.9\pm 1.6$    & $5.8\pm 2.8$\\[2pt]
$B^0\to f_0(980)K^0$, $f_0\to \pi^+\pi^-$ &$\mathcal{A}$     & $-0.14\pm0.22$&$0.01\pm 0.10$  & $0.0004\pm 0.0010$\\[2pt]
                                          &$\mathcal{S}$     & $-0.39\pm0.26$&$-0.63\pm 0.09$ & $-0.77\pm 0.0004$\\
\hline\\[-11pt]
 \raisebox{-1.5ex}[0pt][0pt]{$B^\pm\to \left(K^+K^-\right)_SK^\pm $}         &$\mathcal{B}$   & $<2.9$~~~~\protect{\cite{bell0412}}  & $1.8\pm 0.4$   & $1.7\pm 0.7$\\[2pt]
                                          &$\mathcal{A}_{CP}$& no data       &$-0.44\pm 0.12$ & $0.29\pm 0.21$\\
\hline\\[-11pt]
                                          &$\mathcal{B}$     & no data       &$1.1\pm 0.3$    &$1.2\pm 0.5$\\
 \raisebox{-1.5ex}[0pt][0pt]{$B^0\to \left(K^+K^-\right)_SK_S^0$}      &$\mathcal{A}$     & $-0.09\pm0.10$&$0.01\pm 0.10$  & $0.001\pm 0.001$\\[2pt]
            &\raisebox{-1.5ex}[0pt][0pt]{$\mathcal{S}$}     &
	    $-0.55\pm0.22$~~~~\protect{\cite{baba0408076}}&\raisebox{-1.5ex}[0pt][0pt]{$-0.64\pm 0.09$} &
	    \raisebox{-1.5ex}[0pt][0pt]{$-0.77\pm 0.0006$}\\
                                          &     & $-0.74\pm0.27$~~~~\protect{\cite{bell0409}}& & \\
\hline\\[-11pt]                                         
                                          &$\mathcal{B}$     & no data       &$1.1\pm 0.3$    &$1.2\pm 0.5$\\
 $B^0\to \left(K^0_SK^0_S\right)_SK_S^0$  &$\mathcal{A}$     & $0.41\pm0.21$ &$0.01\pm 0.10$  & $0.001\pm 0.001$\\
                                          &$\mathcal{S}$     & $-0.26\pm0.34$&$-0.64\pm 0.09$ & $-0.77\pm 0.0006$\\
\hline

\end{tabular}
\label{tab:results}
\end{center}
\end{table*}

Within our model one finds the same asymmetries for the 
$B^0 \to K^0_S K^0_S K^0_S$ decay as for the 
$B^0 \to (K^+ K^-)_S K^0_S$ process 
provided that the \mbox{$S$-wave} is dominant in the production of the 
$K^0_SK^0_S$ pairs and their effective masses are not large.
The new data of Belle~\cite{bell0411} 
($\mathcal A = 0.54 \pm 0.34 \pm 0.09$, $\mathcal S=1.26 \pm 0.68 \pm 0.20$) 
and BaBar~\cite{baba05} 
($\mathcal{A}=0.34^{+0.25}_{-0.28}\pm 0.05$, 
$\mathcal S=-0.71^{+0.38}_{-0.32}\pm 0.04$) agree for 
$\mathcal{A}$ and disagree for $\mathcal S$. 
Our results for $\mathcal S$ are close to the BaBar value, however the 
experimental errors of both collaborations are still too large to make 
a definite conclusion.

\section{Summary and outlook}
We have analyzed the 
charged and neutral three-body $B$ meson decays 
into  the $\pi^+\pi^-K$, $K^+K^-K$ and $K^0_SK^0_SK^0_S$ 
systems. 
Our primary aim is a construction of 
decay amplitudes including the
$\pi^+\pi^-$ final state interactions in a~rather wide range of 
effective masses
between the $\pi\pi$ threshold and 1.2 GeV.
The model is based on a factorization approximation with some QCD 
corrections and contains the dominant charming penguin terms. 
Using a single amplitude, we are able to 
describe simultaneously the production of two
scalar-isoscalar resonances $f_0(600)$ and $f_0(980)$. 
The $B$ decay amplitudes 
to the $\pi\pi K$ and $K\overline K K$ states
are connected as the coupling between the $\pi\pi$ and $K\overline K$ 
channels above $1$ GeV is incorporated in our model.
No adjustable free parameters, like arbitrary phase factors between
contributions of different resonances, are needed. 
We have obtained a good agreement with most of the recent BaBar and 
Belle data for the $\pi\pi$ effective mass distributions, the
branching ratios and the time-dependent $CP$ violating asymmetries.
Our results are summarized in table~\ref{tab:results}. 
These numbers depend only weakly on the choice of the renormalization 
scale, the corresponding quark masses $m_b$ and $m_s$, and the value of
$F_0^{B\to K}(0)$. The changes 
are smaller than the errors in the determination of the charming penguin 
amplitudes.

If we use the charming penguin amplitudes determined 
in~\cite{groo03} then the direct $CP$
violation asymmetry in the charged $B$ decays to $\pi\pi K$ is strongly negative
($\sim-0.5$). 
However, for the charming penguin amplitudes taken 
from~\cite{ciuc04} this asymmetry is positive
($\sim+0.2$). 
A similar difference is found for the 
$B^\pm \to K^+K^- K^\pm$ reactions.
Future independent measurements of the $B^\pm \to \pi^+\pi^-K^\pm$ and 
$B^\pm \to K^+K^-K^\pm$ asymmetries will be crucial for a decisive 
test of the phase of the
long distance contributions.
Let us stress the importance of charming penguin amplitudes. 
If we omit them then the average branching ratio for the 
$B^0 \to \pi^+\pi^-K^0_S$ 
decay is too small by a factor of 18 and that of the 
$B^\pm \to \pi^+\pi^-K^\pm$ mode by a factor 
of 4. 
Even  an {\it ad hoc} adjustment of the 
constant $\chi$  to fit the
experimental charged $B$~decays does not allow one to 
explain the neutral $B$ decays. 
However, when the charming penguin amplitudes 
are included, we get a good agreement for both channels 
for $\chi$ values close to the estimation 
based on the $f_0(980)$ properties. 

The model presented in this paper can be extended to 
larger effective $\pi \pi$ mass range, in particular to  
the range where the $f_0(1500)$ is 
important. One can also 
include the final state interactions between one of the
pions and the kaon in the $B$ or $D$ decays to the $\pi\pi K$ 
system. Especially
interesting are the $\pi^-K^+$ or $\pi^+ K^-$ 
subsystems where scalar and vector
resonances can play an important role.

\section*{Acknowledgements}
We acknowledge important discussions and correspondence with 
A. H\"ocker, useful information from J. Chauveau, L. Hinz and  
S. Laplace, enlightening comments from  W. N. Cottingham and good 
advices from B. El-Bennich, J. P. B. C. de Melo and P.~\.Zenczykowski.
One of us (L. L.) would like to thank Maria R\'o\.za\'nska for numerous 
helpful discussions
and Alexei Garmash for useful E-mail communication.

This work has been performed in the framework of the IN2P3-Polish 
Laboratories Convention (project No 99-97).

 \end{document}